\documentclass[twocolumn,floatfix,showpacs,preprintnumbers,amsmath,amssymb,prb]{revtex4}
\usepackage{graphicx}
\usepackage{dcolumn}
\usepackage{color}
\usepackage{bm}
\begin{document}

\title{Emergence of noncollinear magnetic ordering in small magnetic clusters: Mn$_n$ and As@Mn$_n$ clusters}
\author{Mukul Kabir,$^{1,2,*}$ D. G. Kanhere$^3$ and Abhijit Mookerjee$^1$}
\affiliation{$^1$Department of Materials Science, S.N. Bose National Center for Basic Sciences, JD Block, Sector III, Salt Lake City, Kolkata 700 098, India.\\ 
$^2$Department of Materials Science and Engineering, Massachusetts Institute of Technology, Cambridge, Massachusetts 02139, USA.\\
$^3$Department of Physics and Center for Modeling and Simulation, University of Pune, Pune - 411 007, India.}

\date{\today}

\begin{abstract}
Using first-principles density functional calculations, we have studied the magnetic ordering in pure Mn$_n$ ($n=$2$-$10, 13, 15, 19)
and As@Mn$_n$ ($n=$1$-$10) clusters. Although, for both pure and doped manganese clusters, there exists many collinear and noncollinear
isomers close in energy, the smaller clusters with $n\leqslant$5 have collinear magnetic ground state and the emergence of noncollinear 
ground states is seen for $n\geqslant$6 clusters. Due to strong $p-d$ hybridization in As@Mn$_n$ clusters, the binding energy is substantially
enhanced and the magnetic moment is reduced compared to the corresponding pure Mn$_n$ clusters.
\end{abstract}
\maketitle

\section{\label{sec:introduction}Introduction}
Noncollinear magnetism, i.e. magnetic ordering where the local magnetic moments are
not parallel or antiparallel to a global direction, exists in a variety of
systems, for example, in topologically frustrated antiferromagnets, in
spiral spin-density waves and spin-spiral states.
These types of 
orderings can be called interatomic noncollinear magnetism since the different 
atomic moments are noncollinear. Generally, noncollinear configurations occur more 
easily in a low symmetry or a disordered magnetic system.\cite{lorentz, liebs}
 Thus small 
clusters, which have less symmetry constraints than the bulk, are likely candidates for noncollinear structures. Indeed, as a consequence of spin-spiral 
ground state of fcc Fe (Refs. 3 and 4) and spin density wave ground 
state of bcc Cr (Ref. 5), the small Fe$_n$ and Cr$_n$ clusters  
were found to have noncollinear magnetic structures.\cite{oda, ivanov, hobbs,kohl,ojeda}

From the structural and magnetic point of view, manganese is one of the 
most complex of 
all metals and has attracted considerable attention. 
The electronic configuration in a Mn atom is 4$s^2$ 3$d^5$ and consequently
Mn atoms have high magnetic moments of 5 $\mu_B$. High 4$s^2$3$d^5 $ $\rightarrow$ 4$s^1$3$d^6$ 
promotion energy means that Mn atoms do not bind strongly when they are brought together to
form a cluster or a bulk solid.
The most stable polymorph, $\alpha-$Mn, has an exotic crystal structure 
containing 58 atoms in a cubic unit cell.  This $\alpha-$Mn exhibits a complex 
antiferromagnetic order below the N\'{e}el temperature of 95 K and is 
nonmagnetic at room temperature.\cite{tebble} The magnetic transition of $\alpha$-Mn
is coupled to a tetragonal distortion. 
Recent density functional calculations indicate a noncollinear magnetic ground state 
for $\alpha-$Mn.\cite{hobbs1} The other polymorph of Mn, known as $\beta-$Mn, 
also has a stable noncollinear configuration even though the ground state is 
weakly ferrimagnetic.\cite{hafner} These results are in good agreement with the
neutron scattering,\cite{neutron} magnetic torque\cite{magtorque}
and nuclear magnetic resonance\cite{NMR} experiments, where the experimental results 
are interpreted in terms of noncollinear antiferromagnetic structure.

As mentioned earlier small Mn clusters, which have less symmetry constraints, 
are likely candidates for the occurrence of noncollinear magnetic structure. However, 
majority of the 
theoretical calculations have been made under the collinear spin 
assumption.\cite{bobadova1, bobadova2, jones, briere, kabir1}  Among them, 
most extensive study has been done by Kabir {\it et al.}, where the structural and
magnetic properties of Mn$_n$ clusters in the size range $n=$2$-$20 have been investigated
using density functional theory.\cite{kabir1} It was found that Mn$_2$, Mn$_3$ and 
Mn$_4$ exhibit ferromagnetic ordering, whereas a magnetic transition to the ferrimagnetic ordering
takes place at $n=$ 5 and the ferrimagnetic state  continues to be the ground state for clusters with $n>$ 5. The 
predicted magnetic moments are in well agreement with the Stern-Gerlach (SG) clusters 
beam experiments.\cite{mark1, mark2}
However very recently,\cite{kabir2} we pointed out that the noncollinear treatment of spins is 
indeed necessary  at 
the size range $n\geqslant$6.  Morisato {\it et al.} studied Mn$_5$ and Mn$_6$ 
clusters and found Mn$_6$ to be the smallest cluster to exhibit noncollinear ground 
state.\cite{morisato}  However,   using local spin density approximation Longo 
{\it et al.} found that all the clusters in the size range $n=$ 3$-$7 have noncollinear ground 
states.\cite{longo} 
Recently, noncollinear magnetic ordering has also been found in small Mn-clusters 
supported on a Cu(111) surface.\cite{anders} In this study however, the cluster 
geometries have been assumed planer and they were placed on a regular 
face centered-cubic Cu 
lattice with experimental lattice parameters. No structural relaxation for either the 
surface or the cluster was considered. 

The present paper is the sequel of our earlier reports.\cite{kabir1, kabir2}.
In this paper, we allow noncollinear ordering of atomic moments and extensively study how 
collinear and noncollinear magnetic ordering evolves in small
Mn$_n$ clusters and how the magnetic ordering  gets perturbed if we dope a single nonmagnetic atom 
(for example an As-atom) into it.
This is important as (III,Mn)V semiconductors show ferromagnetism 
and the effect of Mn-clustering is being 
debated for a long time.\cite{kabir2, sullivan, zajac, nature, mahadevan, schil, xu} 
Moreover, the noncollinear ferromagnetism 
is common in (III,Mn)V semiconductors.\cite{potashnik1, potashnik2, john1, john2, fiete} 
Experimental study by Potashnik {\it et al.}\cite{potashnik1, potashnik2}
indicated that the Curie temperature, the ground state saturation magnetization
$M (T=0)$ and the shape of the $M(T)$ curve all depend upon the temperature and
annealing. These can be explained only if the noncollinearity in the localized Mn
magnetic moments is considered in the ground state.\cite{john1, john2, fiete}
From this point of view, it will be very interesting to investigate whether  small 
As@Mn$_n$ clusters do show noncollinearity or not.

\section{\label{sec:method}Method}
In the theoretical approaches to noncollinear magnetism,\cite{barth, sandratskii, kubler1, kubler2}
the wave functions are described by two-component
spinors, $\Psi(\mathbf{r}) \equiv (\Psi^{\alpha}(\mathbf{r}),\Psi^{\beta}
(\mathbf{r}))$, where $\alpha$ and $\beta$ are the spin indices. The density
matrix is defined as, 
\begin{equation}
\rho_{\alpha\beta} = \sum_i f_i \Psi^{\alpha}_i(\mathbf{r}) 
\Psi^{\beta\ast}_i(\mathbf{r}),
\end{equation}
where $f_i$ is the occupation number of the single-particle state.\\  
The charge $n(\mathbf{r})$ and magnetization $\vec{m}(\mathbf{r})$ parts of 
the density matrix can be extracted by expanding in terms of the Pauli spin
matrices $\sigma_k$ $(k=x, y, z)$,
\begin{equation}
\rho(\mathbf{r}) = \frac{1}{2}\left[n(\mathbf{r})\mathbf{1} + 
\sum_k m_k(\mathbf{r})\sigma_k\right],
\end{equation}
where $m_k$ is the Cartesian components of $\vec{m}(\mathbf{r})$. The 
exchange-correlation potential, $v_{xc} = v_0(\mathbf{r})\mathbf{1} +
\vec{\xi}(\mathbf{r}) \cdot \vec{\sigma}$, contains nonmagnetic and magnetic
parts. The nonmagnetic part $v_0$ and $|\vec{\xi}|$ are given as a function
of $n$ and $|\vec{m}|$ in the same way that is done in the case of collinear
magnetism, but here in addition $\vec{\xi}(\mathbf{r})$ is always parallel
to $\vec{m}(\mathbf{r})$. In this scheme, the individual eigenstates can have
different spin quantization directions and furthermore, the spin quantization
axis of the each state can vary with position. 

Calculations have been performed using density functional theory (DFT) within the pseudopotential
plane wave method. The projector augmented wave method\cite{kresse1} 
has been used and for the spin-polarized
gradient approximation to the exchange-correlation energy we used the Perdew-Burke-Ernzerhof
functional,\cite{perdew} as implemented in the Vienna {\it ab-initio} 
Simulation Package.\cite{kresse2}
The 3$d$, 4$s$ for Mn and 4$s$, 4$p$ orbitals for As are treated as valence states. 
The spinor wave functions are expanded in a plane wave basis set with the kinetic energy 
cutoff 337.3 eV. 
We have ensured that the cutoff is sufficient by varying it up to 600 eV and found that
the energy difference between ferromagnetic and antiferromagnetic states  converges within 
$\sim$ 1 meV for Mn$_2$ and As@Mn$_2$.
Calculations have been carried out at the $\Gamma$-point. 
We adopted periodic boundary conditions and described the cluster within a large 
simple cubic supercell such that the periodic images are separated by at least 12 \r{A}
of vacuum. This essentially ensures negligible interactions between the periodic images.
Symmetry unrestricted optimizations were performed using 
quasi-Newtonian and conjugate gradient methods until all the force components become 
less than 0.005 eV/\r{A}. For a particular sized Mn$_n$ and As@Mn$_n$ cluster, several different 
initial structures were studied to
ensure that the globally optimized geometry does not correspond to a local minima,
and to be extensive, both noncollinear and collinear magnetic structures have been 
considered separately. Moreover, we explicitly considered all possible spin multiplicities 
for the collinear case. 

For a collinear spin cluster all the spins are parallel ($0^0$) or antiparallel ($180^0$) to each other. 
On the other hand, the angle between any two moments could be anything in between $0^0$ and $180^0$ for a noncollinear case and the
deviation from $0^0$ or $180^0$ is termed ``degree of noncollinearity". The average degree of noncollinearity
for a particular cluster can be defined as,
\begin{equation}
\theta = \frac{\sum_{i,(<j)} \left| \Theta - \theta_{ij} \right|}{\sum_k^{N-1} (N -k)},
\end{equation}
where $ij$ runs for all independent spin pairs and $\sum_k^{N-1}(N-k)$ is the total number of such independent 
spin(atom) pairs with $N$ being the number of atoms in the cluster. $\Theta$ is either $0^0$ or $180^0$
and $\theta_{ij}$ is the angle between $i$-th and $j$-th moment.
 
\section{\label{sec:results}Results and Discussion}
\subsection{\label{sec:pure}Pure Mn$_n$ clusters: Collinear vs noncollinear ordering}

\begin{figure*}[t]
\begin{tabular}{cc}
\begin{minipage}{0.65\textwidth}
\rotatebox{0}{\resizebox{11cm}{14cm}{\includegraphics{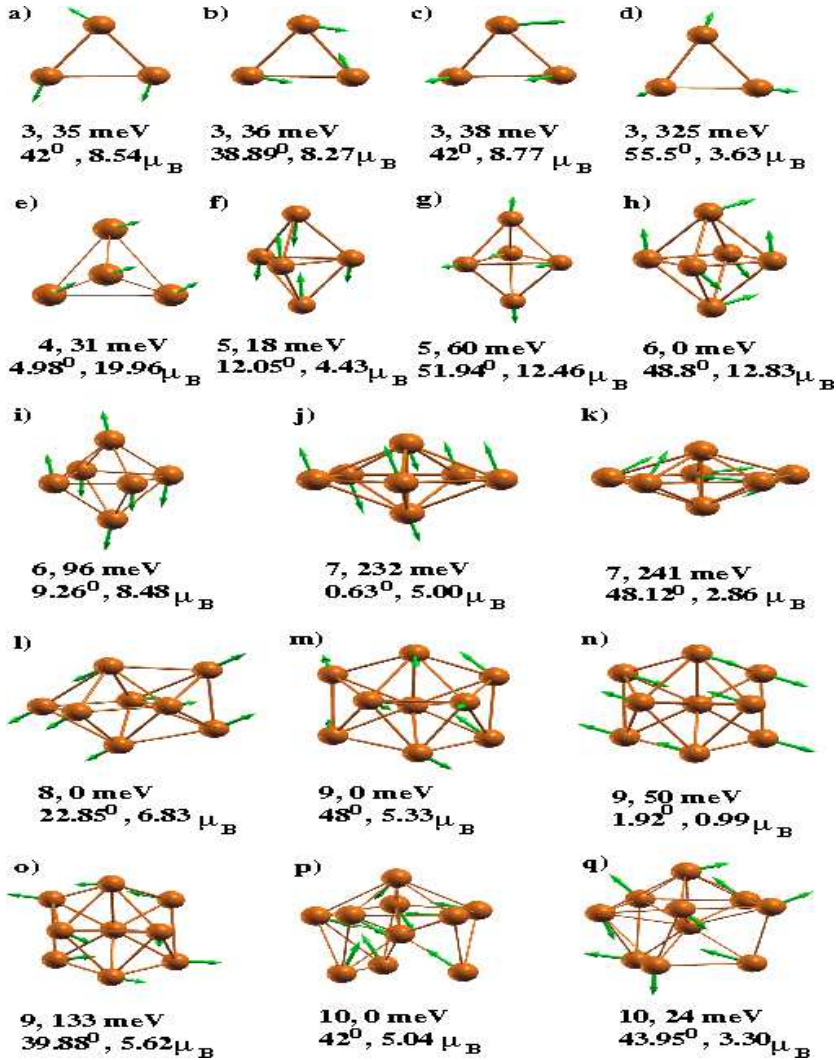}}}
\end{minipage}  & 
\begin{minipage}{0.30\textwidth}
\caption{\label{fig:composite}(color online) Optimal noncollinear structures for Mn$_n$ clusters in the size
range $n=$3-10. The first line gives the number of atoms $n$ in the cluster and the 
energy relative to the corresponding ground state, $\Delta E$, (meV),  whereas, the second line represents
the average degree of noncollinearity ($\theta$ in degree) and the corresponding total magnetic moment ($\mu_B$). 
The optimal collinear structures are not shown here rather we refer to Kabir {\it et al.}.\cite{kabir1}}
\end{minipage}
\end{tabular}
\end{figure*}

\begin{figure}[!t]
\rotatebox{0}{\resizebox{8.0cm}{14.0cm}{\includegraphics{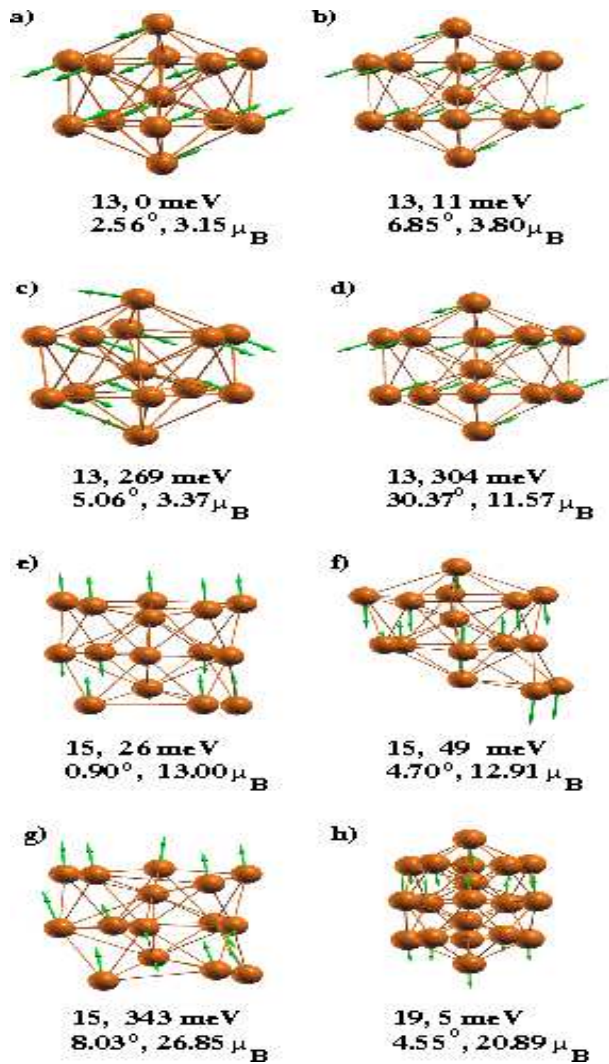}}}
\caption{\label{fig:mn13}(color online) Optimal noncollinear magnetic ordering for Mn$_{13}$, Mn$_{15}$ and Mn$_{19}$.}
\end{figure}

In an earlier work, we have studied the magnetic ordering of pure 
Mn$_n$ ($n \leqslant 20$) clusters within the collinear atomic moment assumption.\cite{kabir1}
Above a certain cluster size ($n$=5), the magnetic ordering is found to be
ferrimagnetic and the calculated magnetic moments were in good agreement\cite{kabir1} with
Stern-Gerlach experiments.\cite{mark1, mark2}
However, the validity of the collinear spin assumption should be checked rigorously
because, in principle, the complex ferrimagnetic ordering and magnetic `frustration'
\cite{bramwell} could definitely lead to the noncollinear ordering in magnetic clusters. 
The type of magnetic ordering along with the total magnetic moment and relative energy difference with the ground state
are given in the Table \ref{tab:table1} and the optimal noncollinear structures are shown 
in Fig.\ref{fig:composite} ($n=$3$-$10) and Fig.\ref{fig:mn13} ($n=$ 13, 15 and 19) for pure Mn$_n$ clusters.

\begin{table}[!t]
\caption{\label{tab:table1} Type of magnetic ordering, average degree of noncollinearity ($\theta$), total magnetic moment
($M_{\rm tot}$) and the relative energy difference ($\Delta$E = E$-$E$_{\rm GS}$) for pure Mn$_n$ clusters for 
$n=$2$-$10, 13, 15 and 19.}
\begin{center}
\begin{tabular}{lllll}
\hline
\hline
Cluster \ \ \ & Magnetism \ \ \ \ \ \ \    & $\theta$ \ \ \ \ \ \ \ \ \ \    & $M_{\rm tot}$ \ \ \ \ \ \ \  & $\triangle$E \\
        &            & ( $^0$)  & ($\mu_B$) &    (meV)       \\
\hline
Mn$_2$   & collinear    & -      &    10                 &  0 \\

Mn$_3$  & collinear     & -      & 15                    &  0 \\
        & noncollinear  & 42     & 8.54                  &  35  \\
        & noncollinear  & 38.89  & 8.27                  &  36   \\
        & noncollinear  & 44     & 8.77                  &  38    \\ 
        & collinear     & -      & 5                     &  46     \\
        & noncollinear  & 55.5   & 3.63                  &  325     \\ 

Mn$_4$  & collinear     & -      & 20                    &  0 \\
        & noncollinear  & 4.98   & 19.96                 &  31 \\
        & collinear     & -      & 10                    &  78  \\

Mn$_5$ & collinear      & -      & 3                     &  0  \\
       & noncollinear   & 12.05  & 4.43                  &  18  \\
       & noncollinear   & 51.94  & 12.46                 &  60   \\
       & collinear      & -      & 13                    &  60    \\
       & collinear      & -      & 5                     &  79     \\

Mn$_6$ & noncollinear   & 48.8   & 12.83                 &  0 \\
       & noncollinear   & 9.26   & 8.48                  &  96 \\  
       & collinear      & -      & 8                     & 123 \\
       & collinear      & -      & 2                     & 140  \\

Mn$_7$ & collinear      & -      & 5                     &  0  \\
       & collinear      & -      & 7                     &  91  \\
       & collinear      & -      & 3                     &  192   \\
       & noncollinear   & 0.63   & 5.00                  &  232   \\
       & noncollinear   & 48.12  & 2.86                  &  241    \\  

Mn$_8$ & noncollinear   & 22.85  & 6.83                  &  0  \\
       & collinear      & -      & 8                     &  170  \\
       & collinear      & -      & 12                    &  170   \\
       & collinear      & -      & 10                    &  200   \\
Mn$_9$ & noncollinear   & 48     & 5.33                  & 0 \\
       & noncollinear   & 1.92   & 0.99                  & 50  \\
       & noncollinear   & 39.88  & 5.62                  & 133 \\ 
       & collinear      & -      & 7                     & 181   \\
Mn$_{10}$& noncollinear & 42     & 5.04                  & 0 \\
         & noncollinear & 43.95  & 3.30                  & 24 \\ 
         & collinear    & -      & 14                    & 81 \\

Mn$_{13}$& noncollinear & 2.56   & 3.15                  & 0  \\
         & collinear    & -      & 3                     & 0   \\
         & noncollinear & 6.85   & 3.80                  & 11   \\
         & collinear    & -      & 7                     & 77    \\ 
         & noncollinear & 5.06   & 3.37                  & 269   \\
         & noncollinear & 30.37  & 11.57                 & 304  \\

Mn$_{15}$& collinear    & -      &  13                   &  0 \\
         & noncollinear & 0.90   &  13.00                & 26 \\
         & collinear    & -      &  5                    &  33 \\
         & noncollinear & 4.70   &  12.91                & 49 \\
         & noncollinear & 8.03   &  26.85                & 343 \\

Mn$_{19}$& collinear    & -      & 21                    & 0   \\
         & noncollinear & 4.55   & 20.89                 & 5   \\
         & collinear    & -      & 19                    & 9 \\
         & collinear    & -      & 9                     & 74 \\       
\hline
\hline
\end{tabular}
\end{center}
\end{table}

The case of triangular Mn$_3$ is very interesting. 
If the triangle is equilateral and the magnetic ordering is antiferromagnetic then 
the third atom doesn't know how to align and, consequently, is in the `frustrated' 
state. This frustration could either be removed by reducing its 
geometrical symmetry or by adopting noncollinear magnetic order or both. 
The ground state is collinear ferromagnetic with 5 $\mu_B$/atom magnetic moment and 
has equilateral triangular geometry. The optimal noncollinear structure is found to be the next
isomer, which lies 35 meV higher in energy. This structure has high noncollinearity
and has a total magnetic moment of 8.54 $\mu_B$ (Fig.\ref{fig:composite}a). The geometrical 
structure for this state is a isosceles
triangle. We also find several noncollinear (Fig.\ref{fig:composite}b-d) and collinear 
(Table \ref{tab:table1})  
magnetic states which lie close in energy.
However, triangular Mn$_3$ cluster prefers noncollinear 
structure when it is placed on a Cu(111) surface, and each moment lie parallel
to the Cu(111) surface and make an equal angle of 120$^0$ to its neighbor.\cite{anders}
For this case, it has also been pointed out that the intra cluster exchange parameter 
is much stronger than that of inter cluster, which means that the magnetic ordering is not affected if 
cluster-cluster distance becomes very small. 
\begin{table}[!t]
\caption{\label{tab:mn6}Magnetic moments ($\mu_B$) and angles ($\theta_{ij}$ in degree) between 
the moments for the noncollinear ground state of Mn$_6$  with 12.83 $\mu_B$ magnetic moment.}
\begin{tabular}{cc}
\begin{minipage}{0.1\textwidth}
\rotatebox{0}{\resizebox{2.8cm}{2.8cm}{\includegraphics{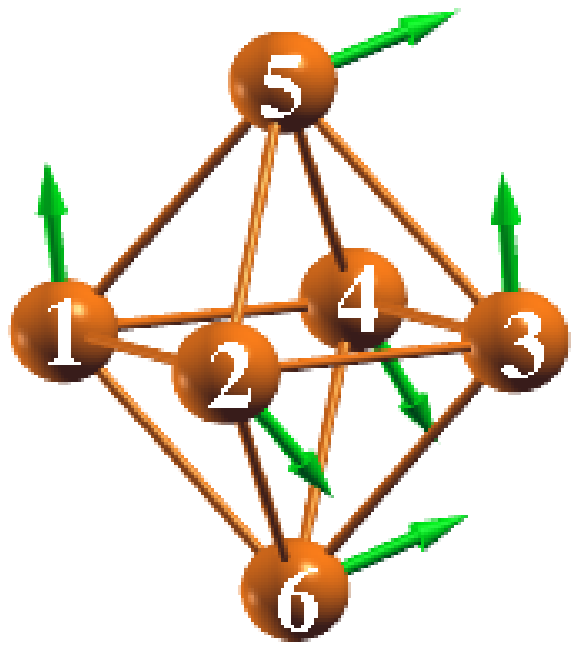}}}
\end{minipage}  & 
\begin{minipage}{0.4\textwidth}
\begin{tabular}{cccccccc}
\hline
$i/j$ & 1 & 2 & 3 & 4 & 5 & 6 & Moment \\
\hline
\hline
1 & -   & 156 & 2   & 146 & 77 & 75 &  3.62\\
2 & 156 & -   & 154 & 9   & 79 & 80 &  3.64 \\
3 & 2   & 154 & -   & 145 & 75 & 74 &  3.62  \\
4 & 146 & 9   & 145 & -   & 70 & 71 &  3.63   \\
5 & 77  & 79  & 75  & 70  & -  & 1  &  3.80    \\
6 & 75  & 80  & 74  & 71  & 1  & -  &  3.80     \\   
\hline
\hline
\end{tabular}
\end{minipage}
\end{tabular}
\end{table}

For the Mn$_4$ cluster, a collinear ferromagnetic ground state with 20 $\mu_B$ magnetic moment  is found to be the
ground state. The optimal noncollinear structure (Fig.\ref{fig:composite}e) lies only 31 meV higher in energy and has comparable magnetic moment. 
In contrast, Longo {\it et al.}\cite{longo} found a noncollinear ground state with very small magnetic
moment, 4.5 $\mu_B$, to be the ground state. It has to be noted that they had used a different level of approximation (local spin density 
approximation (LDA)) for the exchange-correlation, which might be the reason for this discrepancy. Although there are no available report of 
magnetic ordering for Mn$_4$ in the gas phase,  Ludwig {\it et al.}\cite{ludwig} observed a 21-line hyperfine pattern in solid silicon, which 
establishes that all the four atoms in Mn$_4$ are equivalent and has a total magnetic moment of 20 $\mu_B$. 
 
The Mn$_5$ cluster has been investigated by many authors but most of
them have restricted themselves to the collinear spin assumption.\cite{bobadova1, bobadova2,
jones, kabir1} However, few attempts
have made to investigate noncollinear magnetic ordering.\cite{kabir2, morisato,longo} 
We find the collinear ferrimagnetic ground state (Fig.\ref{fig:composite}f) lies slightly lower (18 meV) in energy than the 
optimal noncollinear state. The average degree of noncollinearity found for this structure is 12$^0$.  
The state next higher in energy is the noncollinear structure (12.46 $\mu_B$, $\theta$ $\sim$ 52$^0$
, shown in Fig.\ref{fig:composite}g) and the degenerate ferrimagnetic collinear structure. These are
followed by another collinear ferrimagnetic structure slightly higher in energy.\cite{kabir1} 

The Mn$_6$ cluster is found to be the smallest cluster, which has noncollinear
magnetic order in the ground state (Fig.\ref{fig:composite}h). 
This magnetic structure has high noncollinearity ($\theta \sim$ 49$^0$, the angles $\theta_{ij}$ between the Mn-moments
are given in Table \ref{tab:mn6}) and its geometric structure
is a distorted octahedron. The magnetic moment is 12.83 $\mu_B$.
Present result agrees with Morisato {\it et al.},\cite{morisato} who found a noncollinear
state with 12.24 $\mu_B$ moment as the ground state.
The isomer next higher in energy by 96 meV is the noncollinear structure with total moment 8.48 $\mu_B$. Next higher energy isomers are four collinear structures with moments 8, 2, 16 and 26 $\mu_B$.\cite{kabir1}

The ground state for Mn$_7$ is collinear with a total magnetic moment of 5 $\mu_B$.  
Next higher in energy are two collinear isomers with a total 7 and 3 $\mu_B$  moments (Table \ref{tab:table1}). 
As discussed in our previous work,\cite{kabir1} this is the plausible reason for the 
large experimental uncertainty observed, 0.72 $\pm$ 0.42 $\mu_B$/atom.\cite{mark1}
The optimal noncollinear structure lies much higher (232 meV) in energy and 
has a magnetic moment which is same as the collinear ground state (Fig.\ref{fig:composite}j). 
However, a noncollinear state has been reported to be the ground state in an 
earlier report,\cite{longo} where the exchange-correlation energy has been approximated within LDA.

These results in the size range ($n\leqslant$7) agree well with Morisato {\it et al.}.\cite{morisato} 
They have adopted same level of theory used
in the present report to study Mn$_5$ and Mn$_6$ clusters and found the Mn$_6$ cluster to be 
the smallest one which shows noncollinear magnetic ordering. On the other hand,
Longo {\it et al.}\cite{longo} studied Mn$_3$$-$Mn$_7$ clusters within some different
level of theory. They have used LDA for the exchange-correlation
energy and found the magnetic ordering to be noncollinear for all the clusters in the size range $n=$ 3$-$7.    

As it has been discussed in our previous report\cite{kabir1} that we have considered many 
different initial structures for Mn$_8$ cluster. However, when we allow noncollinearity
among the atomic moments, we find a noncollinear bi-capped octahedral structure
with magnetic moment 6.83 $\mu_B$ to be the ground state (Fig.\ref{fig:composite}l). 
The moment of the noncollinear ground state,  0.85 $\mu_B$/atom,
is very close to the experimental value of 1.04 $\pm$ 0.14 $\mu_B$/atom.\cite{mark2}
This noncollinear state is followed by 
two degenerate collinear magnetic structures and both of them lie 170 meV higher
in energy compared to the ground state. 

A noncollinear centered antiprism with 5.33 $\mu_B$ moment (Fig.\ref{fig:composite}m) is found to 
be the ground state for Mn$_9$ cluster. This structure has very high noncollinearity, $\theta \sim$ 48$^0$.
We find another lees noncollinear state with $\theta$ being 1.92$^0$, which has very tiny magnetic moment 
is the first isomer (Fig.\ref{fig:composite}). This structure lies 50 meV higher in energy compared to the ground state.
The next isomer is also noncollinear in nature
with comparable magnetic moment with the ground state. This structure lies 133 meV higher in energy (Fig.\ref{fig:composite}o). 
The optimal collinear structure lies much, 181 meV, higher in energy and has a  7 $\mu_B$ magnetic moment.    

In our earlier work,\cite{kabir1}, we had found that for the Mn$_{10}$ cluster there existed 
four collinear magnetic structures within a very small energy range of $\sim$ 10 meV.
There we had treated the atomic moments 
collinearly. 
In the present case, we find that a noncollinear magnetic structure (Fig.\ref{fig:composite}p) with magnetic moment 
5.04 $\mu_B$ is lower in energy by 81 meV than the previously found optimal collinear structure. 
This structure is highly noncollinear with $\theta$ being 42$^0$. All these structures, whatever collinear 
or noncollinear, has a pentagonal ring and can be seen as incomplete 13-atom icosahedra.  

In an earlier work\cite{kabir1} for the Mn$_{13}$ cluster, we have investigated icosahedral, cuboctahedral and a hexagonal closed packed 
structures within the
collinear atomic moment  assumption and found an icosahedral structure to be the ground state, with a total magnetic 
moment of 3 $\mu_B$. 
The optimal hexagonal closed packed and cuboctahedral structures were found to have higher magnetic moments, 
9 and 11 $\mu_B$, respectively, and they lay much higher in energy  (0.89 and 1.12 eV, respectively)
compared to the ground state. Here we 
investigate only the icosahedral structure and find a noncollinear magnetic structure (Fig.\ref{fig:mn13}a) with 3.15 $\mu_B$ moment 
to be degenerate with the previously found collinear ground state. The average $\theta$ is found to be small (2.5$^0$) for 
this structure. Another noncollinear structure is nearly degenerate (11 meV) to the ground state with $\theta$ being 6.8$^0$ (Fig.\ref{fig:mn13}b). 
This structure has comparable magnetic moment. 
Another two noncollinear structures (Fig.\ref{fig:mn13}c-d) are found and they 
lie 269 and 304 meV higher, respectively.

Within the collinear atomic moment assumption,\cite{kabir1} we found two different competing icosahedral structures, 
5,1,5,1,3 and 1,5,1,5,1,2 staking, for the Mn$_{15}$ cluster. Five isomers were found within $\sim$ 60 meV  of energy.
The ground state had a magnetic moment of 0.87 $\mu_B$/atom, whereas the experimentally measured value is rather high, 
1.66 $\pm$ 0.02 $\mu_B$/atom.\cite{mark1, mark2} No isomer with comparable magnetic moment was found within the
collinear moment assumption.\cite{kabir1}
At this point it would be interesting to investigate if this large discrepancy 
between the experimental and theoretically predicted magnetic 
moment is originated from collinear assumption. However, the discrepancy do 
not improve when we relax the collinear spin assumption and treat the atomic moments 
noncollinearly (Table \ref{tab:table1}). Two noncollinear structures (Fig.\ref{fig:mn13}e,f) 
lie very close in energy with the previously found collinear ground state. These two structures are of first and second kind 
of icosahedra with $\theta =$ 0.9$^0$ and 4.7$^0$, respectively. However, a structure
of the first kind is found to have a magnetic moment of 1.79 $\mu_B$/atom, which is close to the 
experimental value (1.66 $\pm$ 0.02 $\mu_B$/atom\cite{mark1, mark2}), but this structure 
lies well above (343 meV) the ground state and the magnetic structure is highly noncollinear (Fig.\ref{fig:mn13}g).   

It has experimentally been found that the magnetic moment of Mn$_{19}$ cluster is relatively
smaller than that of its neighboring clusters.\cite{mark1, mark2} Within the collinear 
moment approximation\cite{kabir1}, a double icosahedral structure 
was found to be the ground state, which had a total moment of 21 $\mu_B$. This is somewhat
larger than the experimentally predicted value.\cite{mark1, mark2} 
Another collinear magnetic structure with 
magnetic moment (9 $\mu_B$), which is close to the experimental value lies only 75 meV higher. 
In the present study, which allows noncollinear arrangement of the atomic moments, we found 
the optimal noncollinear structure (Fig.\ref{fig:mn13}h) with low degree of 
noncollinearity to be nearly equienergitic (Table \ref{tab:table1}) with the optimal collinear state. 

\subsection{\label{sec:doping}As@Mn$_n$ clusters: Effect of doping}
\begin{figure*}[!t]
\rotatebox{0}{\resizebox{14.0cm}{14.0cm}{\includegraphics{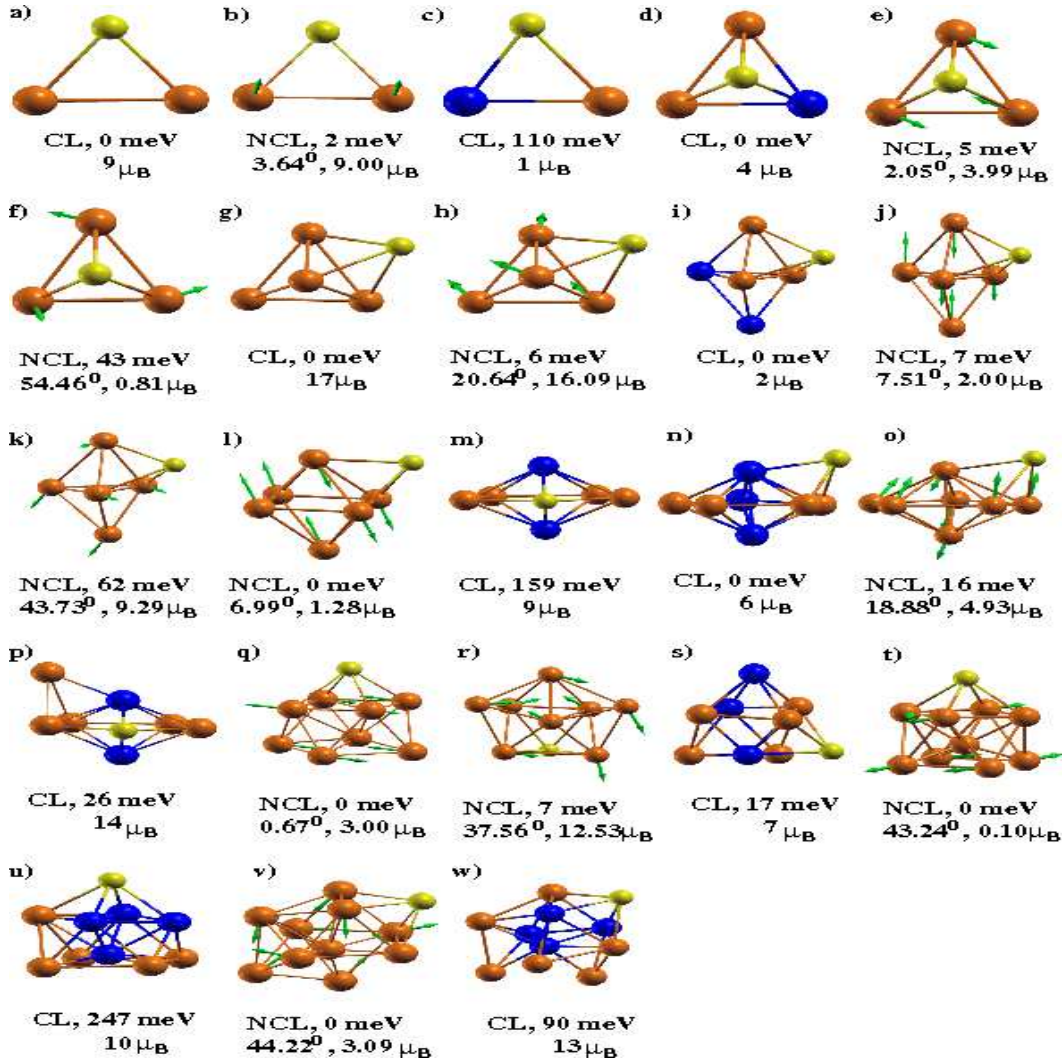}}}
\caption{\label{fig:MnAs_compo}(color online) The ground state and closely lying isomers of As@Mn$_n$ clusters.
The first line indicates the nature of magnetic ordering (collinear (CL) or noncollinear (NCL)) and the relative energy to the corresponding ground
state (meV) and the second line indicates the average $\theta$ (in degree) for NCL case and the total magnetic moment ($\mu_B$). For the 
collinear cases, orange(blue) refers the positive(negative) Mn moment. Yellow atom refers the As-atom for all the
structures.}
\end{figure*}

It would now be interesting to investigate the effect of doping (for example with a single nonmagnetic atom) on the
geometric and magnetic structures of Mn$_n$ clusters. We shall dope a
single nonmagnetic As-atom into an existing Mn$_n$ cluster and shall simultaneously optimize both the geometry 
and magnetic ordering. 
The determination of the true ground state is a delicate task as the number of atoms in the cluster increases.
This was true for the pure Mn$_n$ clusters, where we found many isomers within very 
small energy window.
The situation is equally complicated for the doped case. The Table \ref{tab:table2}
shows the magnetic ordering of the low lying states for As@Mn$_n$ clusters in the size range $n$=1$-$10. The relaxed structures 
are shown in Fig.\ref{fig:MnAs_compo}.
We had discussed\cite{kabir2} that 
the single nonmagnetic As doping does not alter the geometry considerably and can be derived from its 
pure counterpart with 
a moderate perturbation. However, the magnetic structure is strongly influenced by doping, which 
will be discussed  in detail.

\begin{table}[!t]
\caption{\label{tab:table2} Type of magnetic ordering, average degree of noncollinearity\cite{thetaAs@Mn} ($\theta$), total magnetic moment
($M_{\rm tot}$) and the relative energy difference ($\Delta$E) for pure As@Mn$_n$ clusters for 
$n=$1---10.}
\begin{center}
\begin{tabular}{lllll}
\hline
\hline
Cluster  \ \ \ \ \ \ \        & Magnetism \ \ \ \ \ \ \   & $\theta$ \ \ \ \ \ \ \ \ \ \ & $M_{\rm tot}$ \ \ \ \ \ \   & $\Delta$E \\
                              &                            & ( $^0$)                      & ($\mu_B$)               &    (meV)       \\
\hline
\hline
As@Mn           & collinear      &  -             &  4            & 0             \\
As@Mn$_2$       & collinear      &  -             &  9            & 0              \\
                & noncollinear   &  3.64          &  9.00         & 2              \\
                & collinear      &  -             &  1            & 111            \\
As@Mn$_3$       & collinear      &  -             &  4            & 0               \\
                & noncollinear   &  2.05          &  3.99         & 5               \\
                & noncollinear   &  4.79          &  3.97         & 8               \\
                & noncollinear   &  54.46         &  0.81         & 43               \\
As@Mn$_4$       & collinear      &  -             &  17           & 0                \\
                & noncollinear   &  20.64         &  16.09        & 6                \\
                & noncollinear   &  17.63         &  16.33        & 18               \\
As@Mn$_5$       & collinear      &  -             &  2            & 0                \\
                & noncollinear   &  7.51          &  2.00         & 7                \\
                & noncollinear   &  43.73         &  9.29         & 62               \\
                & noncollinear   &  44.26         &  10.08        & 68               \\
As@Mn$_6$       & noncollinear   &  6.99          &  1.28         & 0                \\
                & collinear      &  -             &  9            & 159              \\
                & noncollinear   &  0.91          &  9.00         & 168              \\
                & noncollinear   &  24.04         &  7.59         & 196              \\
As@Mn$_7$       & collinear      &  -             &  6            & 0                \\
                & noncollinear   &  18.88         &  4.93         & 16               \\
                & collinear      &  -             &  14           & 26               \\
                & noncollinear   &  48.77         &  6.28         & 66               \\
                & noncollinear   &  7.59          &  21.83        & 104              \\
                & noncollinear   &  44.59         &  8.62         & 108              \\
As@Mn$_8$       & noncollinear   &  0.67          &  3.00         & 0                \\
                & noncollinear   &  37.56         &  12.53        & 7                \\
                & collinear      &  -             &  7            & 17               \\
                & noncollinear   &  23.45         &  10.84        & 46                \\
As@Mn$_9$       & noncollinear   &  43.24         &  0.10         & 0                 \\
                & collinear      &  -             &  10           & 247               \\
                & noncollinear   &  8.01          &  3.88         & 353               \\
As@Mn$_{10}$    & noncollinear   &  44.22         &  3.09         & 0                 \\
                & noncollinear   &  22.52         &  1.37         & 36                \\
                & collinear      &  -             &  13           & 90                \\
\hline
\hline
\end{tabular}
\end{center}
\end{table}

The AsMn dimer has much higher binding energy of 1.12 eV/atom and much shorter bond length (2.21 \AA) than those of 
Mn$_2$   dimer: 0.52 eV/atom and 2.58 \AA, respectively.\cite{kabir1}
We have repeated our calculations of AsMn dimer including the Mn 3$p$ as valence electrons 
and obtained an optimized bond length of 2.22 \AA \ and binding energy of 1.08 eV/atom, with the same total magnetic moment, which
confirms that the inner Mn-3$p$ electrons contributes insignificantly to the bonding.

The Mn-Mn coupling is found to be ferromagnetic (Fig.\ref{fig:MnAs_compo}a) for As@Mn$_2$ cluster, which has a total magnetic moment of 9 $\mu_B$.
The Mn-Mn distance in this collinear As@Mn$_2$ is same with the pure Mn$_2$ dimer. However, this collinear ground state is nearly degenerate with a noncollinear magnetic structure (Fig.\ref{fig:MnAs_compo}b).
Another collinear structure with antiferromagnetic Mn-Mn coupling (Fig.\ref{fig:MnAs_compo}c) is found to be the next isomer which lies 111 meV higher
than that of the ground state.

The pure Mn$_3$ cluster was found to be ferromagnetic with large 15 $\mu_B$ magnetic moment and is collinear
in nature. The single As-atom doping reduces its magnetic moment considerably to 4 $\mu_B$ for As@Mn$_3$ cluster in its ground state. 
This structure is also collinear in nature, which is shown 
in Fig.\ref{fig:MnAs_compo}d. 
However, as in the pure case, there exist 
several isomers with different magnetic structures (Table \ref{tab:table2} and Fig.\ref{fig:MnAs_compo}e, f). 
The next three isomers are all noncollinear and they lie close in energy, 5, 8 and 43 meV higher, respectively.  

The optimal collinear structure with 20 $\mu_B$ magnetic moment was found to have 31 meV less energy than the optimal noncollinear structure for 
the pure tetramer. On the other hand for As@Mn$_4$ cluster, the optimal noncollinear structure (Fig.\ref{fig:MnAs_compo}h) is found to be 
nearly degenerate with the collinear ferromagnetic ground state (Fig.\ref{fig:MnAs_compo}g). The next isomer (Table \ref{tab:table2}) is also 
noncollinear in nature and all these three structures have comparable magnetic moments. 
  
The ground state of the As@Mn$_5$ cluster is found to be collinear (Fig.\ref{fig:MnAs_compo}i), which is nearly 
degenerate with the optimal noncollinear structure (Fig.\ref{fig:MnAs_compo}j). Both of these structures have equal magnetic
moments, which are particularly small. The next two isomers are noncollinear with high 
noncollinearity and have comparatively large magnetic moments (Table \ref{tab:table2}). 

Similar to the pure Mn$_6$ cluster, the As@Mn$_6$ is found to be the smallest doped cluster, which show noncollinear magnetism (Fig.\ref{fig:MnAs_compo}l) in its
ground state. This noncollinear structure lies 159 meV lower than the next isomer, 
which is collinear (Fig.\ref{fig:MnAs_compo}m). The noncollinear ground state has small magnetic moment (1.28 $\mu_B$) 
compared to the optimal collinear (9 $\mu_B$) state. The noncollinear ground state has a As-capped Mn-octahedral geometry,
whereas the next (collinear) isomer is a pentagonal bi-pyramid where the As-atom sits in the pentagonal ring.
Another noncollinear structure lies 196 meV higher 
which has large noncollinearity (24.04$^0$) and comparable magnetic moment with the collinear structure (see Table \ref{tab:table2}).

The ground state of the As@Mn$_7$ cluster is collinear (Fig.\ref{fig:MnAs_compo}n) with a total magnetic moment of 6 $\mu_B$. The 
next isomer is noncollinear in nature with 4.94 $\mu_B$ magnetic moment (Fig.\ref{fig:MnAs_compo}o), which lies sightly higher 
(16 meV) in energy. However, for pure Mn$_7$ cluster, the energy difference between the collinear
ground state and the optimal noncollinear state is large, 232 meV (Table \ref{tab:table1}); i.e. the single As doping reduces the energy 
difference between the collinear ground state and the optimal noncollinear state. 
Moreover, there exist several collinear and noncollinear structures (Table \ref{tab:table2}), which
are close in energy.

In an early work,\cite{kabir2} we found a collinear ground state with a magnetic moment of 3 $\mu_B$ for As@Mn$_8$ cluster. However, 
after a more rigorous scan\cite{refinedscan} of the potential energy surface we found few different magnetic states (Table \ref{tab:table2}), 
which are lower in energy than the previously reported one. Among them a noncollinear state with very 
low noncollinearity (Fig.\ref{fig:MnAs_compo}q) is found to be the ground state. The next isomer is also noncollinear (Fig.\ref{fig:MnAs_compo}r), 
which is followed by a collinear state (Fig.\ref{fig:MnAs_compo}s) with 7 $\mu_B$ magnetic moment. 

The ground state of As@Mn$_9$ cluster is noncollinear with high noncollinearity, 43.24$^0$.
However, the magnetic moment of this structure is nearly zero, 0.1 $\mu_B$ (Fig.\ref{fig:MnAs_compo}t), which is
much smaller than that of the pure Mn$_9$ cluster in its ground state. The next
isomer is collinear in nature (Fig.\ref{fig:MnAs_compo}u) and has comparatively large magnetic moment, 10 $\mu_B$. 
However, this collinear structure lies much higher, 249 meV, in energy. 

We have found several isomers for the As@Mn$_{10}$ cluster (Table \ref{tab:table2}). Among all a noncollinear magnetic 
structure (Fig.\ref{fig:MnAs_compo}v) with high noncollinearity (45.22$^0$) is found to be the ground state. This structure has a total
magnetic moment of 3.09 $\mu_B$. The next isomer is also noncollinear which lies only 36 meV higher in 
energy. The optimal collinear structure have comparatively high magnetic moment and lies 90 meV higher (Fig.\ref{fig:MnAs_compo}w). 

\subsection{Binding energy}

\begin{figure}[!b]
\rotatebox{270}{\resizebox{6.0cm}{8.5cm}{\includegraphics{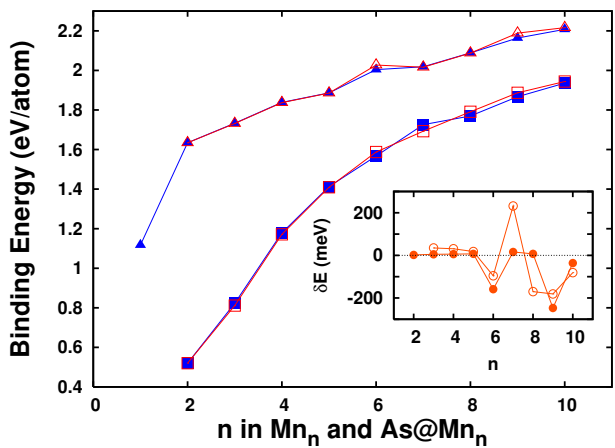}}}
\caption{\label{fig:be} (color online) Plot of binding energy for optimal collinear and noncollinear 
configurations as a function of Mn atoms ($n$) in pure Mn$_n$ and As@Mn$_n$ clusters. 
Binding energy is defined as, BE(Mn$_n$) = $-$[ E(Mn$_n$) $-$ $n$ E(Mn)]/$n$ for pure Mn$_n$
clusters and BE(As@Mn$_n$) = $-$[ E(As@Mn$_n$) $-$ $n$ E(Mn) $-$ E(As)]/($n$+1) for As@Mn$_n$
clusters, where E(Mn$_n$) and E(As@Mn$_n$) are the total energies of pure Mn$_n$ and As@Mn$_n$
clusters, respectively. The $\square$($\blacksquare$) represents the optimal noncollinear(collinear)
structures for pure Mn$_n$ clusters, whereas, $\vartriangle$($\blacktriangle$) represents the optimal
noncollinear(collinear) structures for As@Mn$_n$ clusters. Inset shows the total energy difference
between the optimal collinear(CL) and noncolliear(NCL) configurations ($\delta$E = -[E$_{\rm CL}$($n$) - E$_{\rm NCL}$($n$)]) as a function of $n$. The $\circ$($\bullet$) represent Mn$_n$(As@Mn$_n$) clusters.}
\end{figure}

Binding energies of the optimal collinear and noncollinear magnetic structures have been shown
in Fig.\ref{fig:be} for both pure Mn$_n$ and As@Mn$_n$ clusters. It has been
understood\cite{kabir1} that due to the lack of hybridization between the filled 4$s$ states and 
the half-filled 3$d$ states and due to high (2.14 eV) 4$s^2$ 3$d^5$ $\rightarrow$ 4$s^1$ 3$d^6$
promotion energy, Mn atoms do not bind strongly when they form clusters or crystals. 
This is manifested through their low binding energy (see Fig.\ref{fig:be}), which is the lowest among all other 3$d$-
transition metal clusters. This is also  
experimentally evidenced through recent photodissociation experiments for cationic 
clusters.\cite{teraski, tono} However, the bonding improves considerably due to doping of 
single nonmagnetic As-atom.\cite{kabir2} When an As-atom is attached to the pure Mn$_n$ clusters, due to 
enhanced hybridization through the As-$p$ electrons, the binding energy of the resultant As@Mn$_n$ 
cluster increases substantially. It is found that the energy gain in adding an As-atom to an 
existing pure Mn$_n$ cluster is much larger than that of adding a Mn-atom to an existing 
As@Mn$_{n-1}$ cluster.\cite{kabir2} This clearly indicates that the As-atom may act as a nucleation
center for Mn-atoms. 

The total energy difference, $\delta$E, between the optimal collinear and noncollinear structure for a 
particular sized cluster is plotted in the inset of Fig.\ref{fig:be} as a function of $n$ for both pure
 and doped clusters. By definition if $\delta$E is positive(negative) the corresponding 
ground state is collinear(noncollinear). For both pure Mn$_n$ and As@Mn$_n$ clusters, the collinear 
states are found to be lower in energy than the corresponding optimal noncollinear states up to five
Mn-atoms in the cluster and in the size range $n \geqslant$ 6 the noncollinear states start to be 
the ground state with the exception for $n=$7 for Mn$_n$ and $n=$7 and 8 for As@Mn$_n$ clusters. However, it 
should be noted here that for the entire size range there exists several collinear/noncollinear 
isomers which are close in energy with the corresponding ground state (see Table \ref{tab:table1} and \ref{tab:table2}).
This establishes the importance to treat the moments noncollinearly.

\subsection{Magnetic moment}

\begin{figure}[!t]
\rotatebox{270}{\resizebox{6.0cm}{8.5cm}{\includegraphics{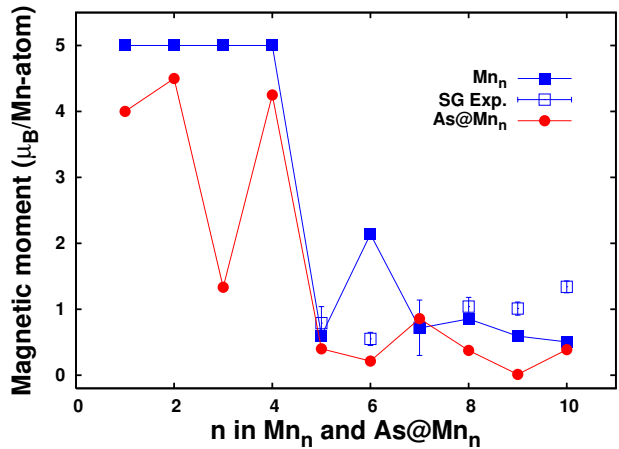}}}
\caption{\label{fig:mag} (color online) Plot of magnetic moment per atom as a function of $n$ for pure Mn$_n$ and
As@Mn$_n$ clusters. Values only for the ground states have only been shown (see Table \ref{tab:table1} and Table \ref{tab:table2} for isomers).
The SG experimental values for pure Mn$_n$ clusters are shown with error bars.}
\end{figure}

The total magnetic moment of pure Mn$_n$ and As@Mn$_n$ clusters of the corresponding ground states and
closely lying isomers are given in Table \ref{tab:table1} and Table \ref{tab:table2}, respectively, and
plotted in Fig.\ref{fig:mag} for the ground states. Pure Mn$_n$ clusters in the size range $n\leqslant$ 4 have
collinear ground state and the Mn-Mn coupling is ferromagnetic. These clusters have a 5 $\mu_B$/atom 
magnetic moment which is the Hund's rule value for isolated Mn-atom. The magnetic moment per atom 
decreases drastically at $n=$ 5 and the magnetic ground states start to be noncollinear for $n\geqslant$ 6.
However, we have found many collinear and noncollinear isomers with varying magnetic moment which are close
in energy (not shown in Fig.\ref{fig:mag}, see for instance Table \ref{tab:table1}). The ground state magnetic 
moments of pure Mn$_n$ clusters are compared with the SG experiment in Fig.\ref{fig:mag}. Calculated 
magnetic moments are in good agreement with the SG experiment. However, the quantitative value differ
for few clusters sizes. For an example, we find a much larger magnetic moment (2.14 $\mu_B$/atom) for
Mn$_6$ cluster compared to the experimental value of 0.55 $\mu_B$/atom.\cite{mark2} However, our 
predicted value agrees with previous DFT calculation.\cite{morisato} This quantitative discrepancy between the calculated and measured magnetic moment may arise
from few reasons: (I) If the true ground state is missed out during the search through the potential energy surface,
which is very much unlikely in the present case as we have extensively considered many different geometric 
and magnetic structures as initial guess. (II)  In every magnetic deflection measurement the magnetic moments
are calculated assuming a model (e.g. superparamagnetic, locked moment, adiabatic magnetization etc.) which 
might affect the outcome. (III) Due to the isomer distribution in the experimental cluster beam.
 
The large moment of pure Mn$_n$ and As@Mn$_n$ clusters arise from localized 3$d$ electrons at Mn-atoms. 
However, the strong $p-d$ hybridization induces a small negative polarization to the As-atom. For example,
this negative polarization is 0.26 $\mu_B$ for AsMn dimer. However, the magnitude of this negative polarization decreases
monotonically to 0.01 $\mu_B$ for As@Mn$_{10}$ cluster. Generally, the total magnetic moment of As@Mn$_n$
cluster is lower than the corresponding pure Mn$_n$ cluster due to $p-d$ hybridization. Similar to their
pure counterparts, the Mn-Mn coupling in As@Mn$_2$ and As@Mn$_4$ is collinear ferromagnetic and the emergence
of noncollinear ground state is seen at $n$=6. Moreover, it has been seen that Mn-Mn exchange coupling in As@Mn$_n$
clusters show anomalous behavior compared to Ruderman-Kittel-Kasuya-Yosida (RKKY) type predictions.\cite{kabir2}
For collinear ground state of As@Mn$_2$ cluster exchange coupling oscillates between positive and negative
with Mn-Mn separation ($r$) favoring the ferromagnetic and antiferromagnetic solutions, respectively, and dies down as 1/$r^3$, which
is a typical RKKY-type behavior. However on the other hand, the average exchange coupling decreases with $n$ and has strong 
environment dependency, which is in contradiction with the RKKY-type prediction.
      
\section{\label{sec:con}Summary and Conclusion}

This study presents a systematic investigation of electronic structure and emergence of noncollinear magnetism
in pure Mn$_n$ and As@Mn$_n$ clusters within gradient-corrected DFT approach.   No considerable structural
change has been found due to noncollinear treatment of atomic moments for both pure and doped manganese 
clusters. Moreover, single As-doping to Mn$_n$ clusters does not give rise any considerable structural 
change, but only a moderate perturbation to their pure counterpart. The pure Mn$_n$ clusters have low 
binding energies, which improves substantially due to single As-doping through strong $p-d$ hybridization.
The ground state of both pure Mn$_n$ and As@Mn$_n$ clusters for $n\leqslant$ 5 is collinear and emergence 
of noncollinear ground states is seen for $n\geqslant$ 6. However, for both kind of clusters, there 
exist many collinear and noncollinear isomers.  Due to strong $p-d$ hybridization the magnetic moments 
of As@Mn$_n$ are smaller compared to their pure counterpart. This also induces a negative polarization to 
the As-atom in As@Mn$_n$ cluster.  Although the results presented here are specific to 
the Mn$_n$ and As@Mn$_n$ clusters, they also contain more general picture: noncollinear magnetic 
ordering is possible in small magnetic clusters.

\acknowledgements 
This work has been done under the Indian Department of Science and Technology Grant No. SR/S2/CAMP-25/2003.


\begin{thebibliography}{99}
\bibitem{refinedscan} After our initial report,\cite{kabir2} we have considered several more initial structures
to confirm the true ground state, for each cluster.  However, we did not find any new structures 
with energy lower than those previously reported ground states except only for As@Mn$_8$ cluster.
\end{thebibliography}
\end{document}